\documentclass[prd,aps,nofootinbib,%preprintnumbers,showpacs,
showkeys]
{revtex4}
\usepackage{graphicx,epsf,amsfonts,amssymb,amsbsy}
\newcommand{\ds}{\displaystyle}
\newcommand{\vev}[1]{\langle#1\rangle}
\newcommand{\mat}{\left ( \begin{array}}
\newcommand{\emat}{\end{array} \right )}
\newcommand{\vect}{\left ( \begin{array}{c}}
\newcommand{\evect}{\end{array} \right )}

\begin{document}

\title{ \bf 
Charged pion condensation and color superconductivity phenomena in 
chirally asymmetric dense quark matter}
\author{%M. M. Gubaeva$^{1}$, 
T. G. Khunjua $^{1}$, K. G. Klimenko $^{2}$, and R. N. Zhokhov $^{2,3}$ }

%\affiliation{$^{1}$ Dubna State University (Protvino branch), 142281 Protvino, Moscow Region, Russia}
\affiliation{$^{1}$ The University of Georgia, GE-0171 Tbilisi, Georgia}
%\affiliation{$^{2)}$ Department of Theoretical Physics, A. Razmadze Mathematical Institute, I. Javakhishvili Tbilisi State University, GE-0177 Tbilisi, Georgia}
%\affiliation{$^{1)}$ A. Razmadze Mathematical Institute, Georgian Academy of Sciences, 380093 Tbilisi, Republic of Georgia}
\affiliation{$^{2}$ State Research Center
of Russian Federation -- Institute for High Energy Physics,
NRC "Kurchatov Institute", 142281 Protvino, Moscow Region, Russia}
%\affiliation{$^{3)}$ University ``Dubna`` (Protvino branch), 142281, Protvino, Moscow Region, Russia}
\affiliation{$^{3}$  Pushkov Institute of Terrestrial Magnetism, Ionosphere and Radiowave Propagation (IZMIRAN),
108840 Troitsk, Moscow, Russia}

\begin{abstract}
In this paper, the question of the influence of color superconductivity (CSC) on the 
formation of a phase with condensation of charged pions in dense chirally asymmetric quark 
matter is studied. We consider it within the framework of the massless NJL model with a 
diquark interaction channel at zero temperature, but in the presence of baryon $\mu_B$, 
isospin $\mu_I$, chiral $\mu_{5}$ and chiral isospin $\mu_{I5}$  chemical potentials. 
It has been shown in the 
mean-field approximation that in the presence of chiral imbalance, when $\mu_{5}\ne 0$ 
and/or $\mu_{I5}\ne 0$, CSC phenomenon does not hinder the generation of charged 
pion condensation in dense quark matter even at largest baryon densities attainable in 
heavy ion collisions experiments and in cores of neutron stars and its mergers. So charged pion condensation in dense quark matter with chiral imbalance predicted earlier is not in a bit suppressed by the presence of CSC, highlighting the resilience of this phase in dense quark matter. 
%It was shown that pion condensation in dense quark matter and color superconductivity phenomena as a rule occupies different regions of phase diagram, so one can say that intriguingly they peacefully coexist and share the phase diagram.
%It shows that phase structure of dense quark matter has a very rich and interesting structure and is very multifaceted in its nature. Encompassing various phenomena in its scope
%This make phase structure of dense quark matter even more rich, multifaceted and interesting in context of, for example, heavy ion collisions experiments terms of various phenomena.
This makes phase structure of dense quark 
matter even more rich, multifaceted and interesting, and shows that pion condensation in dense quark matter is viable phenomenon to be explored, for example, in intermediate energy heavy ion collisions experiments.
%This makes pion condensation in dense quark possible to play a role and be detected in real physical 
%Intriguingly, pion condensation in dense quark matter and color superconductivity phenomena can coexist, inhabiting different regions of phase diagram. This shows the multifaceted nature of QCD phase diagram and the intricate interplay between different phenomena.

%The findings of this study underscore the significance of chiral imbalance as a catalyst for CPC. Moreover, they demonstrate that CPC is not entirely suppressed by the presence of CSC, highlighting the resilience of this phase in dense quark matter.

%The unveiled phase structure of dense quark matter, with CPC and CSC intertwining and coexisting, unveils the complexity and richness of this exotic state of matter. This intricate interplay between the phases underscores the need for further investigations to fully comprehend the behavior of dense quark matter and its implications for astrophysical phenomena such as neutron star interiors and heavy ion collisions.

\end{abstract}

\maketitle
%\draft
%\large
\maketitle

\section{Introduction}

Quantum chromodynamics (QCD) is the fundamental theory of strong interactions, which govern the behavior of quarks and gluons, the fundamental constituents of matter. Dense quark matter, hypothetical state of matter which could exist in the cores of compact stars or arise in heavy-ion collisions, is a subject of intense research. However, perturbative QCD methods are not applicable to this regime due to the large value of the strong coupling constant, and as a result effective QCD-like models, such as the Nambu--Jona-Lasinio (NJL) model, 
are often used to study its properties \cite{njl,buballa,buballa2,zhokhov1,zhokhov2}.

The NJL model 
is based on the assumption of four-fermion interactions among quarks, which effectively 
account for the effects of quark-gluon interactions at low energies. In order to investigate
the properties of quark matter with nonzero baryon $n_B$ and nonzero isospin $n_I$ densities
(which are characteristic properties of the dense matter found in neutron stars), one 
extends this model and incorporates the terms with baryon $\mu_B$ and isospin $\mu_I$ chemical 
potentials. These chemical potentials are nothing more than quantities that are thermodynamically 
conjugate to baryon and isospin densities, respectively. And in terms of 
$\mu_B$ and $\mu_I$ the task of studying the phase structure of low-energy quark matter is 
significantly simplified \cite{son,he,ak,ekkz,Mammarella:2015pxa,Andersen:2018nzq,Ayala,kyll}.

In addition to isospin asymmetry (when $n_I\ne 0$), recently the possible emergence of chiral
asymmetry (or chiral imbalance) in quark matter, and its manifestations has been drawing more attention. Usually chiral asymmetry, i. e.
unequal densities $n_L$ and $n_R$ of all left- and all right-handed quarks of a system, is characterized
by the quantity $n_5$, called the chiral density, $n_5\equiv n_R-n_L$. Chiral imbalance 
can be generated dynamically due to the Adler-Bell-Jackiw anomaly and the interaction of 
quarks with gauge field configurations, called sphalerons, 
in high temperature environment, for example in heavy ion collisions 
\cite{andrianov,cao,braguta,braguta2,Farias}. Let us note that in 
the most general case, the chiral densities of $u$ and $d$ quarks separately, namely $n_{u5} = n_{uR} 
- n_{uL}$ and $n_{d5} = n_{dR} - n_{dL}$, are not necessarily equal. In this case, an 
additional chiral isospin density emerges, denoted by $n_{I5} = n_{u5} - n_{d5}$. This 
chiral isospin density reflects the unequal densities of left- and right-handed $u$ and $d$ 
quarks, contributing to the overall chiral asymmetry of quark matter.
The presence of $n_{I5}$ density further enriches the phase structure of 
dense quark matter, encompassing the effects of both chiral and isospin imbalances. 
It is also important to note that, as it was discussed in Ref. \cite{Khunjua:2019lbv}, the 
chiral imbalance can be created not only in the medium with high temperature, but strong magnetic fields can also induce the emergence of chiral and chiral isospin 
densities even in cold (with zero or not so large values of temperatures) dense quark
matter, for example, in magnetars, i.e. neutron stars, which have exceptionally strong 
magnetic fields.

The presence of baryon density ($n_B\ne 0$), isospin asymmetry ($n_I\ne 0$), as well as two
types of chiral imbalance, when $n_5\ne 0$ and/or $n_{I5}\ne 0$, introduces four additional 
parameters beyond temperature that govern the behavior of dense quark matter. 
As a consequence, baryonic matter composed of $u$ and $d$ quarks can be characterized by 
four additional physical parameters: baryon chemical potential $\mu_B$, isospin chemical 
potential $\mu_I$, chiral chemical potential $\mu_5$, and chiral isospin chemical potential 
$\mu_{I5}$. Then the most general
$(\mu_B,\mu_I,\mu_5,\mu_{I5})$-phase diagram of the simplest NJL model  with massless 
quarks, in which the phenomenon of color superconductivity was not taken into account, has
been explored in the mean-field approximation in Refs. \cite{kkz18,kkz18-2}. (The predictions of this simplest NJL model are equivalent to considering physical processes within the framework of QCD, but only
in the region of rather small energies and/or baryon densities, i.e. at $\mu_B\lesssim 1$ GeV.) This investigation demonstrated that only two nontrivial phases are allowed for dense quark
matter in this scenario: (i) chiral symmetry breaking (CSB) phase and (ii) charged pion 
condensation (PC) one. Furthermore, it was established that in this case there is duality between CSB and charged PC phenomena. A more detailed investigation of the impact of this dual symmetry on the phase structure of dense quark matter has been performed in the framework of the NJL model in Refs. \cite{zhokhov1, zhokhov2,kkz18,kkz18-2,khunjua,khunjua2,khkz}. These investigations have unveiled a curious capability of duality to streamline and simplify the study of the phase structure of quark matter, particularly when it is exposed to multiple external conditions, such as different chemical potentials.

At rather high baryon densities, i.e. at $\mu_B\gtrsim1$ GeV,  quarks are packed so closely together that they can pair up and form bound states called diquarks. These diquarks can then condense, leading to a new state of matter called color superconductivity (CSC). This phase can be obtained in the framework of NJL model with four-fermion interaction terms responsible for the diquark interaction channel (see, e.g., the reviews \cite{alford}).
Let us note that in the papers \cite{zhokhov1, zhokhov2,kkz18,kkz18-2,khunjua,khunjua2,khkz} duality has been established in the framework of simplest NJL model, which does not contain diquark interaction channel, i.e. under assumption that color superconductivity is absent. 
Then it was shown in terms of the NJL model approximation that the duality between CSB and charged PC is respected by color superconductivity phenomenon as well \cite{CSC}.

Recall that the idea of PC phenomenon in baryon matter has been discussed for the first time in the early seventies in terms of nuclear matter, i.e. in the case when baryon density is much less than in dense quark matter. 
It was also understood that pion condensation phenomenon could be a significant factor in the physical processes of dense nuclear matter \cite{pionnuclear} (see also, for example, the reviews \cite{pionnuclearVoskresensky}). Pion condensation in nuclear matter still draws attention \cite{Voskresensky:2022gts, Yasuda, VoskresenskyPion, Dohi:2021lbu}. Though let us note that there are some subtleties, for example, the s-wave charged PC is considered highly unlikely to be realized in nuclear matter \cite{Ohnishi:2008ng}, but it was argued that p-wave charged PC is possible \cite{EricsonWeise}. However, quite recently the s-wave pion condensation in the isospin-symmetric nuclear matter was 
revisited in terms of several models in \cite{Voskresensky:2022gts}. Then charged 
pion condensation in quark matter with isospin imbalance has been considered thoroughly 
in \cite{son,he,ak,ekkz,Mammarella:2015pxa,Andersen:2018nzq,kyll} and, in particular, it was 
shown that charged PC appears when isospin chemical potential $\mu_I$ is greater than pion mass. 
The research was performed in terms of different effective models, lattice QCD, chiral perturbation theory, etc.
The pion condensation phenomenon got into spotlight recently concerning various physical systems such as \cite{Kuznietsov:2021lax,Brandt:2018bwq, Andersen:2018nzq,  Stashko:2023gnn} and neutron star mergers 
\cite{Vijayan:2023qrt}. The possibility of pion condensation in the early Universe 
and its implications for gravitational wave signatures has been considered in Refs.
\cite{Vovchenko:2020crk, Cao:2021gfk, Wygas:2018otj}. Moreover, it was shown that rotation of the  media could lead to charged PC phenomenon as well \cite{Liu:2017spl, Cao:2019ctl}.  

As for quark matter, the charged PC phenomenon  has been discussed in \cite{son,he,ak,ekkz,kyll}, where it 
was shown that charged pion condensate could be generated in quark matter with nonzero baryon 
density and isospin imbalance. So charged PC has been predicted at nonzero 
baryon density (below, to emphasize this fact, instead of PC we will sometimes use the notation 
PC$_d$ for the phenomenon, or phase, with pion condensation in quark medium with a nonzero 
baryon density). Though later it became clear that in the physical point the region occupied
by this phase is rather small and its prediction is not robust, and additionally it vanishes
when one requires that the medium is in beta equilibrium and electrically neutral \cite{kyll}, i.e. in the 
case interesting in context of matter in cores of neutron stars. Later there have been shown
that there exist several conditions that could lead to charged pion condensation in dense 
quark matter. There are a lot of details discussed, e.g., in \cite{zhokhov1,zhokhov2}. One of these 
parameters is chiral imbalance in the medium and, as it has been shown in the recent papers \cite{khkz,khunjua2}, 
it is a factor that hugely promotes charged PC phenomenon in dense quark matter even under neutron star conditions. In these papers, the prediction of charged PC$_d$ phase of chirally asymmetric quark matter was made at a rather wide range of baryon chemical  
potential $\mu_B$. However, it must be especially emphasized that in all the papers 
concerning charged PC the possibility of the color superconductivity phenomenon and 
quark-quark condensation was not taken into account,  which significantly undermined confidence in the results of these papers, especially when $\mu_B>1$ GeV. First of all, it was done due to simplicity reasons. Surely, there were already a lot of condensates and chemical  
potentials in the consideration beside CSC, and it was reasonable as a first step. Further, although 
one knows that CSC phenomenon takes place at large baryon density and it is probably a
dominating phase in that region of QCD phase diagram, there was an expectation that at 
least not all regions of charged PC$_d$ phase would be replaced by color superconductivity. Although there were absolutely no solid grounds to guarantee this, %there is only one argument for charged pion condensation survival -- it is a rather large chunk of charged PC$_d$ phase that is generated in a wide range of baryon densities in the presence of chiral imbalance.
%Let us note once again that 
ignoring color superconductivity in Refs. \cite{zhokhov1,zhokhov2,son,he,ak,ekkz,kyll} 
was made for reasons of simplicity and easier considerations %clearer prediction 
of the charged PC$_d$ phase in chirally imbalanced media. 
%In fact, %there was very little reason for this, 
And taking into account color superconductivity could easily change 
all the results of such an analysis in the region of high baryon densities. So omitting the role of color superconductivity leads to an incomplete understanding of PC$_d$ phase generation phenomenon.
So it is reasonable to revise the generation of charged pion condensation in dense chirally asymmetric quark 
matter, taking into account such an important phenomenon as color superconductivity at large baryon densities. 
It is this problem that is the subject of this work, in which, for this purpose, we study the phase structure of the NJL model with an additional diquark interaction channel and in the presence of several chemical potentials, $\mu_B,\mu_I,\mu_5$ and $\mu_{I5}$.
We could go forward a bit and say that it turned out that CSC does not spoil the prediction 
of charged PC generation in dense quark matter. And surprisingly enough CSC phase in a way 
shies away from charged PC$_d$ phase and they realize in different regions of chemical potentials.
The chemical  
potential space is so huge that charged pion condensation in dense quark matter and color 
superconductivity phenomenon has its own niche in the phase diagram and does not bother 
each other. That shows that the bigger the space of chemical  
potentials, i.e. the more diverse conditions the quark matter has, the more interesting 
and replete the phase diagram is.

The paper is organized as follows. In Sec. II a (3+1)-dimensional NJL model with two massless quark flavors ($u$ and 
$d$ quarks) that includes four kinds of chemical potentials, $\mu_B,\mu_I,\mu_{I5},\mu_{5}$, are introduced. 
In addition to usual quark-antiquark channels, the model contains the diquark interaction one and is intended to 
describe the phenomenon of color superconductivity in a dense quark medium. Furthermore, the symmetries of the model 
are discussed and its thermodynamic potential (TDP) is presented in the mean-field approximation.
In Sec. III the phase structure of this NJL model is investigated in the chiral limit. 
In Sec. IV summary and conclusions are given.

\section{The model and its thermodynamic potential}

Our investigation employs NJL type model with two quark
flavors with Lagrangian containing the interaction of quarks both in the
quark--antiquark and scalar diquark channels,
\begin{eqnarray}
 L=\bar q\Big [\gamma^\nu i\partial_\nu-
m\Big ]q+ G\Big [(\bar qq)^2+
(\bar qi\gamma^5\vec\tau q)^2\Big ]+H\sum_{A=2,5,7}
[\overline{ q^c}i\gamma^5\tau_2\lambda_{A}q]
[\bar qi\gamma^5\tau_2\lambda_{A} q^c].
\label{1}
\end{eqnarray}
Here, $q$ is a 
flavor doublet of $u$ and $d$ quarks, color triplet as well as a
four-component Dirac spinor (the corresponding indexes are omitted in Eq. (\ref{1})). The charge-conjugated spinor $q^c$ is defined as $q^c = C \bar{q}^T$, where $C = i \gamma^2 \gamma^0$ is the charge conjugation matrix. Current (bare) mass of the $u$ and $d$ quarks is denoted by $m$, Pauli matrices $\tau_a$ act in flavor space, while Gell-Mann matrices $\lambda_A$ act in color space.

The Lagrangian (\ref{1}) is invariant with respect to transformations from the color SU(3)$_c$ and baryon U(1)$_B$ groups. At $m = 0$, it is also invariant under the chiral SU(2)$_L \times$ SU(2)$_R$ group. However, when $m \neq 0$, this symmetry group is reduced to a diagonal isospin subgroup SU(2)$_I$ with generators $I_k=\tau_k/2$, where $k=1,2,3$.
The electric charge $Q$ and baryon charge $B$ are also conserved quantities in this model. The electric charge is given by $Q = I_3 + B/2$, where $I_3$ is the third component of the isospin operator, and $B$ is the baryon charge operator. These quantities are unit matrices in color space, but in flavor space they are $Q = {\rm diag}(2/3,-1/3)$, $I_3 = {\rm diag}(1/2,-1/2)$, and $B = {\rm diag}(1/3,1/3)$.
%If the Lagrangian is obtained from the QCD one-gluon exchange approximation, the ratio of coupling constants $H/G$ is expected to be close to 0.75 \cite{buballa,alford}. 
In the absence of more rigorous theoretical or experimental prescriptions
we will employ the following value for diquark coupling $H\approx 0.75G$. That could be obtained from QCD in one-gluon exchange approximation and using Fierz transformation in color current interaction \cite{buballa,alford}.

To effectively utilize the model (1) for investigating the properties of dense quark matter, it is necessary to augment Lagrangian (1) with terms that incorporate chemical potentials,
\begin{eqnarray}
L_{dense}=L+\bar q{\cal M}\gamma^0 q\equiv L+\bar q\left [\frac{\mu_B}{3}+\frac{\mu_I}2\tau_3+\frac{\mu_{I5}}2\gamma^5\tau_3+
\mu_5\gamma^5\right ]\gamma^0 q,
  \label{2}
\end{eqnarray}
where the chemical potential matrix ${\cal M}$ encompasses four distinct chemical potential terms, denoted by $\mu_B$, $\mu_I$, $\mu_5$, and $\mu_{I5}$, that allow for the description of quark matter with non-zero baryon, isospin, chiral, and chiral-isospin densities, respectively.
In the case of all nonzero chemical potentials in Eq. (\ref{2}), $SU(2)_I$ at $m\ne 0$ is not the symmetry group of $L_{dense}$. In this case, due to the presence of the $\mu_{I}$ term, the Lagrangian (\ref{2}) instead exhibits symmetry under the flavor $U(1)_{I_3}$ group, represented by $q\rightarrow\exp (\mathrm{i}\alpha\tau_3/2)q$. Nonetheless, in the chiral limit ($m=0$), an additional symmetry arises, $U(1)_{AI_3}:q\rightarrow\exp (\mathrm{i}\alpha\gamma^5\tau_3)q$.

In order to investigate the phase diagram of the system (\ref{2}), we need to derive its thermodynamic potential, for example, within the mean-field approximation. To achieve this, let us consider the linearized version, denoted as ${\cal L}$, of Lagrangian (\ref{2}) that involves auxiliary bosonic fields $\sigma(x)$, $\pi_a(x)$ and $\Delta_A(x)$,
\begin{eqnarray}
{\cal L}\ds &=&\bar q\Big [\gamma^\nu i\partial_\nu
+{\cal M}\gamma^0
 -\sigma - m -i\gamma^5\vec\pi\vec\tau\Big ]q
 -\frac{1}{4G}\Big [\sigma^2+\pi_a^2\Big ]
 \nonumber\\ &-&\frac1{4H}\Delta^{*}_{A}\Delta_{A}-
 \frac{\Delta^{*}_{A}}{2}[\overline{q^c}i\gamma^5\tau_2\lambda_{A} q]
-\frac{\Delta_{A'}}{2}[\bar q i\gamma^5\tau_2\lambda_{A'}q^c].
\label{3}
\end{eqnarray}
In Eq. (\ref{3}) summation over repeated indices $a=1,2,3$ and $A,A'=2,5,7$ is assumed implicitly.
It is straightforward to demonstrate the equivalence of Lagrangians (\ref{2}) and (\ref{3}) by employing the equations of motion for auxiliary  bosonic fields, which are given by 
\begin{eqnarray}
\sigma (x)=-2G(\bar qq),~~\pi_a(x)=-2G(\bar qi\gamma^5\tau_a q),~~
\Delta_{A}(x)\!\!&=&\!\!-2H(\overline{q^c}i\gamma^5\tau_2\lambda_{A}q),~~
\Delta^{*}_{A}(x)=-2H(\bar qi\gamma^5\tau_2\lambda_{A} q^c).
\label{4}
\end{eqnarray}
As it is evident from (\ref{4}), the mesonic fields $\sigma(x)$ and $\pi_a(x)$ are real, i. e. satisfying the condition $(\sigma(x))^\dagger = \sigma(x)$ and $(\pi_a(x))^\dagger = \pi_a(x)$, where the superscript $\dagger$ denotes hermitian conjugation. In contrast, the diquark fields $\Delta_A(x)$ are complex scalars, satisfying $(\Delta_A(x))^\dagger = \Delta_A^*(x)$. 
The real fields $\sigma(x)$ and $\pi_a(x)$ form color singlets, while the scalar diquarks $\Delta_A(x)$ constitute a color antitriplet $\bar{3}_c$ under the group SU(3)$_c$. 
Notably, the auxiliary bosonic field $\pi_3(x)$ corresponds to the real $\pi^0(x)$ meson, while the physical $\pi^\pm(x)$ mesons are formed by combining other composite fields from equation (\ref{4}), $\pi^\pm(x) = (\pi_1(x) \mp i\pi_2(x))/\sqrt{2}$. 
There occurs a spontaneous breakdown of the color symmetry in the model (\ref{2}) when any of the scalar diquark fields acquires a nonzero expectation value in the ground state, $\vev{\Delta_{A}(x)}\ne 0$.

In the mean-field approximation the effective action ${\cal S}{\rm {eff}}(\sigma,\pi_a,\Delta_{A},\Delta^{*}_{A'})$ of the model (\ref{2}) can be expressed using the path integral over quark fields:
\begin{eqnarray}
\exp(i {\cal S}_{\rm {eff}}(\sigma,\pi_a,\Delta_{A},
\Delta^{*}_{A'}))=
  N'\int[d\bar q][dq]\exp\Bigl(i\int {\cal L}\,d^4 x\Bigr),\label{10}
\end{eqnarray}
where
\begin{eqnarray}
&&{\cal S}_{\rm {eff}}
(\sigma,\pi_a,\Delta_{A},\Delta^{*}_{A'})
=-\int d^4x\left [\frac{\sigma^2+\pi^2_a}{4G}+
\frac{\Delta_{A}\Delta^{*}_{A}}{4H}\right ]+
\tilde {\cal S}_{\rm {eff}},
\label{11}
\end{eqnarray}
and $N'$ is a normalization constant.
The quark contribution to the effective action, represented by the term $\tilde {\cal S}_{\rm {eff}}$ in (\ref{11}), is expressed as follows:
\begin{eqnarray}
\exp(i\tilde {\cal S}_{\rm {eff}})=N'\int [d\bar
q][dq]\exp\Bigl(i\int\Big [\bar q Dq-
 \frac{\Delta^{*}_{A}}{2}[\overline{q^c}i\gamma^5\tau_2\lambda_{A} q]
-\frac{\Delta_{A'}}{2}[\bar q i\gamma^5\tau_2\lambda_{A'}q^c]\Big ]d^4 x\Bigr),%\nonumber\\
%&=&N'\int [d\bar
%q][dq]\exp\Bigl(\frac{i}{2}\int\Big [\bar
%qD^+q+\overline{ q^c}D^-q^c-\bar qK q^c-\overline{q^c}K^{*}q\Big ]d^4 x\Bigr).
\label{12}
\end{eqnarray}
The notation has been used, where $1\!\!{\rm I}_{3_c}$ is the unit operator in the tree-dimensional color space,
\begin{eqnarray}
D=\big(\gamma^\nu i\partial_\nu
+{\cal M}\gamma^0
 -\sigma (x) - m -i\gamma^5\vec\pi (x)\vec\tau\big)\cdot 1\!\!{\rm I}_{3_c}, 
\label{13}
\end{eqnarray}
Starting from Eqs. (\ref{11}) and (\ref{12}), one can define the thermodynamic potential (TDP) $\Omega(\sigma,\pi_a, \Delta_{A}, \Delta^{*}_{A'})$ of the model
(\ref{2}).  Indeed, this quantity is defined by the following relation,
\begin{equation}
{\cal S}_{\rm {eff}}~\bigg
|_{~\sigma,\pi_a,\Delta_{A},\Delta^{*}_{A'}=\rm {const}}
=-\Omega(\sigma,\pi_a,\Delta_{A},\Delta^{*}_{A'})\int d^4x.
\label{17}
\end{equation}
The ground state expectation values of the bosonic fields (\ref{4}), i.e. the quantities $\vev{\sigma(x)},~\vev{\pi_a(x)},~\vev {\Delta_{A}(x)},~\vev{\Delta^{*}_{A'}(x)}$, are solutions obtained by minimizing the thermodynamic potential $\Omega(\sigma,\pi_a,\Delta_{A},\Delta^{*}_{A'})$. They satisfy the gap equations for the TDP (usually are the coordinates of the global minimum point (GMP) of $\Omega$ as a function of $\sigma,\pi_a,\Delta_{A},\Delta^{*}_{A'}$)
\begin{eqnarray}
\frac{\partial\Omega}{\partial\pi_a}=0,~~~~~
\frac{\partial\Omega}{\partial\sigma}=0,~~~~~
\frac{\partial\Omega}{\partial\Delta_{A}}=0,~~~~~
\frac{\partial\Omega}{\partial\Delta^{*}_{A'}}=0.
\label{18}
\end{eqnarray}
At nonzero bare quark mass, $m\neq 0$, the chiral condensate $\vev{\sigma(x)}$ is always nonzero \footnote{See, e.g., the gap equations (13) and (14) of Ref. \cite{Ebert}, which 
at $m\ne 0$ have single solution $\vev{\sigma(x)}\ne 0$ and $\vev{\pi_3(x)} = 0$.}. In contrast, in the chiral limit, $m=0$, it is possible that chiral condensate vanishes, leading to the chiral symmetric phase. 
However, at $m = 0$ the model has a more complex phase structure, because its phase diagram can contain both a region with $\vev{\sigma(x)}=0$, which corresponds to a chiral symmetrical phase, and a region with $\vev{\sigma(x)}\ne 0$, and in the last case chiral symmetry is broken spontaneously.
In addition,  both at $m=0$ and $m\neq 0$, the global minimum point of the TDP  with $\vev{\pi^\pm(x)}\neq 0$ is allowed, which  corresponds to the charged pion condensation (PC) phase. Finally, recall that if one of the scalar diquark fields acquires a nonzero ground state expectation value, $\vev{\Delta_{A}(x)}\neq 0$, a new phase emerges with spontaneously broken color $SU(3)_c$ symmetry, known as the color superconducting (CSC) phase.

The Lagrangian (\ref{2}) and the effective action (\ref{11}) are both invariant under the color $SU(3)_c$ group. 
This invariance implies that the TDP (\ref{17}) depends on the combination $\Delta_2\Delta^*_2+\Delta_5\Delta^*_5+\Delta_7\Delta^*_7 = \Delta^2$, where $\Delta$ is a real quantity. In the chiral limit, where the model exhibits a $U_{I_3}(1)\times U_{AI_3}(1)$ invariance, the TDP (\ref{17}) effectively depends on the combinations $\sigma^2+\pi_3^2$ and $\pi_1^2+\pi_2^2$, in addition to $\Delta$. It is reasonable to assume that $\pi_2=\pi_3=0$, this allows us to study the TDP as a function of only three variables: $\sigma$, $\pi_1$, and $\Delta$. Therefore, $\Omega(\sigma,\pi_1,\Delta)$ can be derived by setting $\Delta_2=\Delta_2^*=\Delta$, $\Delta_5=\Delta_7=0$, and $\pi_2=\pi_3=0$ in Eqs. (\ref{11}) and (\ref{12}). 
For simplicity, we also assume that all auxiliary bosonic fields are independent of the space-time coordinates.

It can be shown that the contribution of the blue $q_b$ quarks in the expression  (\ref{12}) is factorized, i.e.
\begin{eqnarray}
\exp(i\tilde {\cal S}_{\rm {eff}})&=&N'\int [d\bar
q_b][dq_b]\exp\Bigl(i\int\Big [\bar q_b D^+q_b\Big ]\Bigr)\nonumber\\
&\times&
\int [d\overline Q][dQ]\exp\Bigl(i\int\Big [\overline Q \big(D^+\cdot 1\!\!{\rm I}_{2_c}\big) Q-
 \frac{\Delta}{2}[\overline{Q^c}i\gamma^5\tau_2\sigma_{2} Q]
-\frac{\Delta}{2}[\overline Q i\gamma^5\tau_2\sigma_{2}Q^c]\Big ]d^4 x\Bigr),
\label{210}
\end{eqnarray}
where
quark fields $q_b$ and $Q$ represents the flavor doublet of blue quarks and the flavor ($u$ and $d$) and color (red and green) quark doublet, respectively.
And operator $1\!\!{\rm I}_{2_c}$ denotes the unit operator in the two-dimensional (of red and green quarks) color space.
Additionally, the notation $D^+$ is used to represent the operator that is explicitly defined below in Eq. (\ref{14}).

After carrying out the functional integrations in Eq. (\ref{210}) (the integration over the blue quarks $q_b$ is straightforward, and the one for quarks $Q$ can be found, e.g., in Appendix of Ref. \cite{CSC}) one can get
\begin{eqnarray}
\exp(i\tilde {\cal S}_{\rm {eff}})&=&N'\det D^+\cdot{\det}^{1/2}(Z),
\label{22}
\end{eqnarray}
where
\begin{equation}
Z=\left (\begin{array}{cc}
D^+\cdot 1\!\!{\rm I}_{2_c}, & -K\\
~-K~~~~ , &D^-\cdot 1\!\!{\rm I}_{2_c}
\end{array}\right ),\label{15}
\end{equation}
and
 \begin{eqnarray}
&&D^+=i\gamma^\nu\partial_\nu- m+{\cal M}\gamma^0-\Sigma,~~~\Sigma=\sigma+ i\gamma^5\pi_1\tau_1,
\nonumber\\
&&D^-=i\gamma^\nu\partial_\nu- m-\gamma^0 {\cal M}-\Sigma,~~~
K=i\Delta\gamma^5\tau_2\sigma_{2}.
\label{14}
\end{eqnarray}
It is important to note that  $Z$-matrix elements in Eqs. (\ref{15}) and (\ref{14}) are operators that act both in the two-dimensional color and flavor spaces, as well as in the four-dimensional spinor and coordinate spaces.
Then, one can get from Eqs. (\ref{11}) and (\ref{22}) the expression for effective action 
\begin{equation}
{\cal S}_{\rm
{eff}}(M,\pi_1,\Delta)
=-\int d^4x\left[\frac{(M-m)^2+\pi^2_1}{4G}+
\frac{\Delta^2}{4H}\right]-\frac i2\ln\det (Z)-i\ln\det (D^+),
\label{16}
\end{equation}
where the gap $M\equiv\sigma+m$. The last term of Eq. (\ref{16}), which does not depend on $\Delta$, was calculated 
in our recent paper \cite{kkz18} (see there Eqs. (16)-(21)),
\begin{eqnarray}
i\ln\det (D^+)=i\int\frac{d^4p}{(2\pi)^4}\ln\Big [\big (\eta^4-2a_+\eta^2+b_+\eta+c_+\big )\big (\eta^4-2a_-\eta^2+b_-\eta+c_-\big )\Big ]
\int d^4x,
\label{07}
\end{eqnarray}
where $\eta=p_0+\mu$, $|\vec p|=\sqrt{p_1^2+p_2^2+p_3^2}$ and
\begin{eqnarray}
a_\pm&=&M^2+\pi_1^2+(|\vec p|\pm\mu_{5})^2+\nu^2+\nu_{5}^2;~~b_\pm=\pm 8(|\vec p|\pm\mu_{5})\nu\nu_{5};\nonumber\\
c_\pm&=&a_\pm^2-4 \nu ^2
\left(M^2+(|\vec p|\pm\mu_{5})^2\right)-4 \nu_{5}^2 \left(\pi_1^2+(|\vec p|\pm\mu_{5})^2\right)-4\nu^{2} \nu_{5}^2
\label{101}
\end{eqnarray}
(we also use in Eqs. (\ref{07}), (\ref{101}) and below the notations $\mu=\mu_B/3$, $\nu=\mu_I/2$ and $\nu_5=\mu_{I5}/2$). The 
next undefined term of Eq. (\ref{16}) is the following
\begin{eqnarray}
\frac i2\ln\det (Z)=i\int\frac{d^4p}{(2\pi)^4}\ln\det L(p)\int d^4x,
\label{20}
\end{eqnarray}
where $L(p)$ is a matrix acting in 4-dimensional spinor space. Its matrix elements
were obtained in our previous work \cite{CSC} (the same matrix $L(p)$ also arises when 
studying the phase structure of the two-color NJL model \cite{2color}). Here we only need 
the eigenvalues $\tilde\lambda_i(p)$ of $L(p)$, where $i=1,...,4$, with the help of which we can find its determinant,
\begin{eqnarray}
\det L(p)=\tilde\lambda_1(p)\tilde\lambda_2(p)\tilde\lambda_3(p)\tilde\lambda_4(p),
\label{21}
\end{eqnarray}
appearing on the right side of equality (\ref{20}).
Then, taking into account the definition (\ref{17}),
one can get the expression for TDP of the model, 
\begin{eqnarray}
\Omega(M,\pi_1,\Delta)
&=&\left[\frac{(M-m)^2+\pi^2_1}{4G}+
\frac{\Delta^2}{4H}\right]+i\int\frac{d^4p}{(2\pi)^4}\ln
\big[\tilde\lambda_1(p)\tilde\lambda_2(p)\tilde\lambda_3(p)\tilde\lambda_4(p)\big]
\nonumber\\
&+&i\int\frac{d^4p}{(2\pi)^4}\ln\Big [\big (\eta^4-2a_+\eta^2+b_+\eta+c_+\big )\big (\eta^4-2a_-\eta^2+b_-\eta+c_-\big )\Big ].
\label{19}
\end{eqnarray}
Here the notations (\ref{101}) are used and
\begin{eqnarray}
&&\widetilde\lambda_{1,2}(p)=\lambda_{1,2}(p)\Big |_{|\vec p|\to|\vec p|-\mu_5},~~\widetilde\lambda_{3,4}(p)=
\lambda_{3,4}(p)\Big |_{|\vec p|\to|\vec p|+\mu_5}
\label{71}
\end{eqnarray}
where
\begin{eqnarray}
&&\lambda_{1,2}(p)=N_1\pm 4\sqrt{K_1},~~\lambda_{3,4}(p)=N_2\pm 4\sqrt{K_2},
\label{56}
\end{eqnarray}
\begin{eqnarray}
\hspace{-1cm}N_2=N_1+16\mu\nu\nu_5|\vec p|,~~K_2=K_1+8\mu\nu\nu_5|\vec p|p_0^4-8\mu\nu\nu_5|\vec p|p_0^2\big (M^2+\pi_1^2+|\Delta|^2
+|\vec p|^2+\mu^2+\nu^2-\nu_5^2\big ),&&
\label{57}
\end{eqnarray}
\begin{eqnarray}
%\hspace{-1cm}
K_1=\nu_5^2p_0^6-p_0^4\Big [2\nu_5^2 \big(|\Delta|^2+\pi_1^2+M^2+|\vec p|^2+\nu^2+\mu^2-\nu_5^2\big)+4\mu\nu\nu_5|\vec p|\Big ]+
p_0^2 \Big\{
\nu_5^6+2\nu_5^4 \big (M^2-|\Delta|^2-\pi_1^2&&~~~~~~~~~~~~~~\nonumber\\
-\nu^2-\mu^2-|\vec p|^2\big )+4\mu^2\nu^2\big(M^2+
|\vec p|^2\big)+4|\vec p|\mu\nu\nu_5\big(|\Delta|^2+\pi_1^2+M^2+|\vec p|^2+\nu^2+\mu^2-\nu_5^2\big)&&\nonumber\\
+\nu_5^2\Big [\big(|\Delta|^2+\pi_1^2+|\vec p|^2+\nu^2+\mu^2\big)^2+2|\vec p|^2M^2+M^4+2M^2\big(|\Delta|^2-\nu^2+\pi_1^2-\mu^2\big)\Big ]\Big\},&&
\label{58}
\end{eqnarray}
\begin{eqnarray}
N_1&=&p_0^4-2p_0^2\Big [|\Delta|^2+\pi_1^2+M^2+|\vec p|^2+\nu^2+\mu^2-3\nu_5^2\Big ]+
\nu_5^4-2\nu_5^2\Big [|\Delta|^2+\pi_1^2+|\vec p|^2+\nu^2+\mu^2-M^2\Big ]\nonumber\\
&-&8\mu\nu\nu_5|\vec p|+\left (|\vec p|^2+M^2+\pi_1^2+|\Delta|^2-\mu^2-\nu^2\right )^2-
4\left (\mu^2\nu^2-\pi_1^2\nu^2-|\Delta|^2\mu^2\right ).
\label{59}
\end{eqnarray}
Previous studies of the phase structure of two-color QCD (see, for example, Ref. \cite{2color}) have yielded exact expressions for the eigenvalues $\tilde\lambda_i(p)$.
Using Eq. (\ref{19}) for the TDP, we are ready now to discuss in the chiral 
limit, i.e. at $m=0$, the phase structure of the NJL model (\ref{1})-(\ref{2}) in the mean-field approximation.

Throughout the paper we use in numerical investigations of the TDP (\ref{19}) 
the soft cut-off 
regularization scheme when $d^4p\equiv dp_0d^3\vec p\to dp_0d^3\vec p f_{\Lambda}(\vec p)$. 
Here the cut-off function is 
\begin{eqnarray}
f_{\Lambda}(\vec p)=\sqrt{\frac{\Lambda^{2N}}{\Lambda^{2N}+|\vec p|^{2N}}}, 
\end{eqnarray}
and the parameter fit used is $G=4.79$ GeV$^{-2}$, $\Lambda=638.8$ MeV and $N=5$.

\section{Charged pion condensation at non-zero baryon density}

As discussed above in the Introduction, the PC phenomenon in dense quark matter has been discussed in various conditions, for example in nuclear matter \cite{pionnuclear,pionnuclearVoskresensky}, in matter with isospin imbalance, in quark/baryonic matter with baryon density and isospin imbalance, then it was shown that charged PC could be generated in quark matter with nonzero baryon density \cite{he}. But this prediction was not robust since later it was shown that  in the physical point (and in the absence of chiral asymmetry) charged PC  occupies only a rather small region of phase diagram and, additionally it vanishes when one considers electrically neutral medium in beta equilibrium, i.e. the matter in cores of neutron stars \cite{kyll}.  Then several scenarios where PC is generated in the dense quark matter has been discussed.  Among them there has been shown that chiral imbalance, when $\mu_5$ and/or $\mu_{I5}$ are nonzero, leads to generation of charged pion condensation in dense quark  matter, even under the neutron stars conditions \cite{khkz,khunjua2}. Pion condensation discussed in terms of nuclear matter takes place at rather low baryon density and it cannot be challenged by CSC phenomenon, since the densities are not high enough. Charged PC predicted in dense quark matter takes place at higher baryon densities, in the realm of color superconductivity phenomenon, and it is different story. However, in all the papers concerning charged PC in dense quark matter the possibility of the CSC phenomenon and quark-quark condensation was not considered just for simplicity. Surely, there were a lot of condensates and chemical potentials in the consideration beside it, and it was reasonable as a first step.
Since at large baryon densities the CSC phenomenon is believed to be an important feature of the QCD phase diagram, it is of great interest to explore the situation in more detail.

\begin{figure}
%----figure 1
\includegraphics[width=0.49\textwidth]{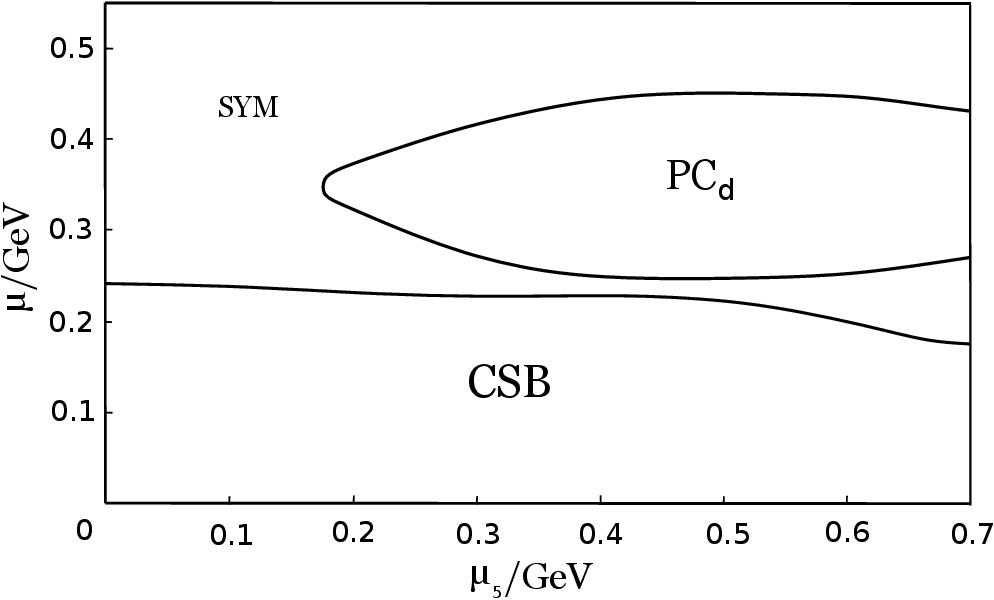}
 \hfill
\includegraphics[width=0.47\textwidth]{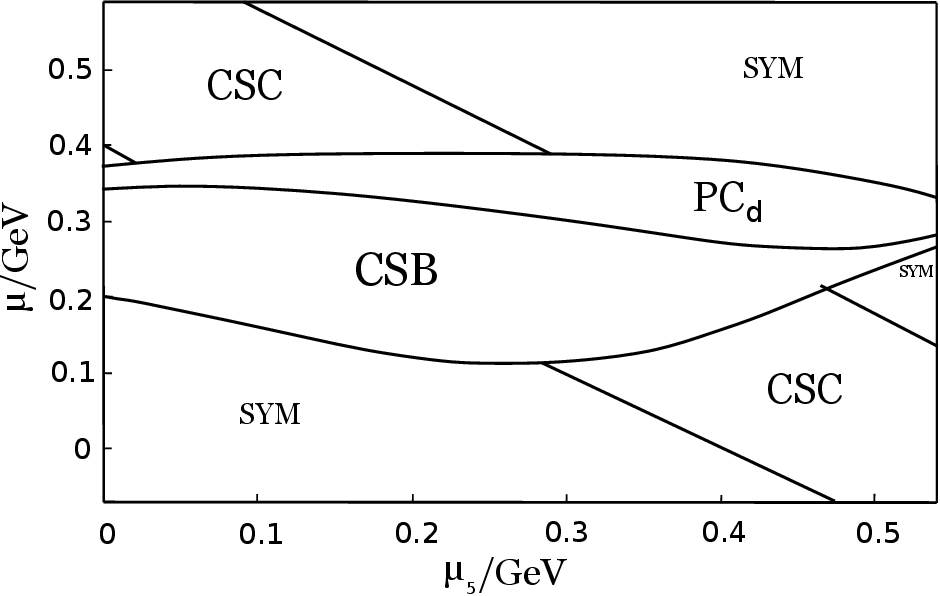}\\
%\label{fig1}
%\parbox[t]{0.45\textwidth}{
\caption{$(\mu_5, \mu)$-phase diagrams of the model at $H=0.75G$. Left panel: The case 
of $\nu_5=-0.35$ GeV and $\nu=0$. 
%}
 %}\hfill
%\parbox[t]{0.45\textwidth}{
%\caption{
Right panel: The case of $\nu_5=0.3$ GeV and $\nu=-0.29$ GeV. Here PC$_d$ denotes
the charged pion condensation phase, CSB and CSC mean respectively the chiral symmetry breaking and color 
superconducting phases, ``sym`` is the symmetric phase. %, where all symmetries are restored.
 } %}
%\label{fig2}
\end{figure}

\begin{figure}
%----figure 2
\includegraphics[width=0.58\textwidth]{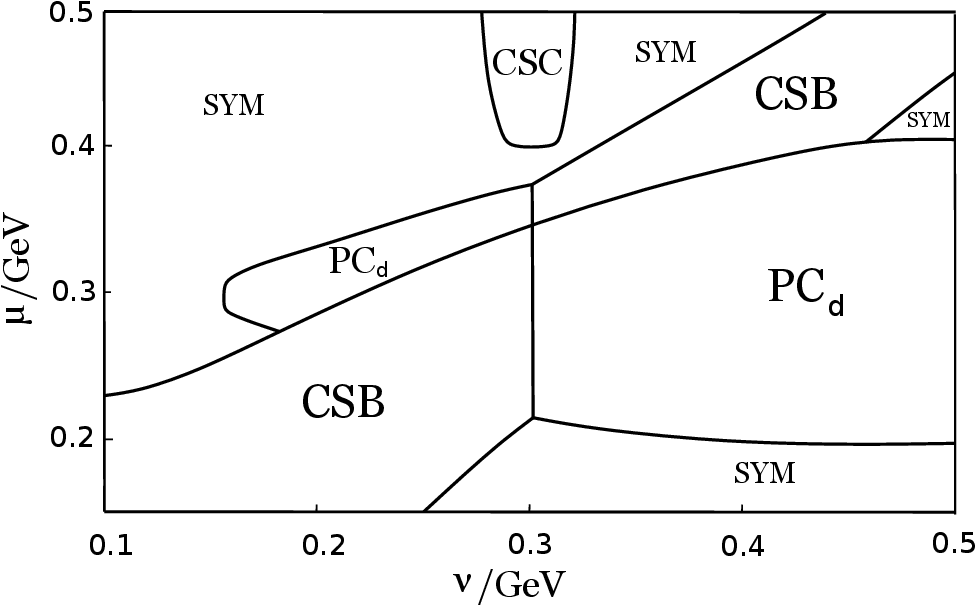}
 \hfill
\includegraphics[width=0.37\textwidth]{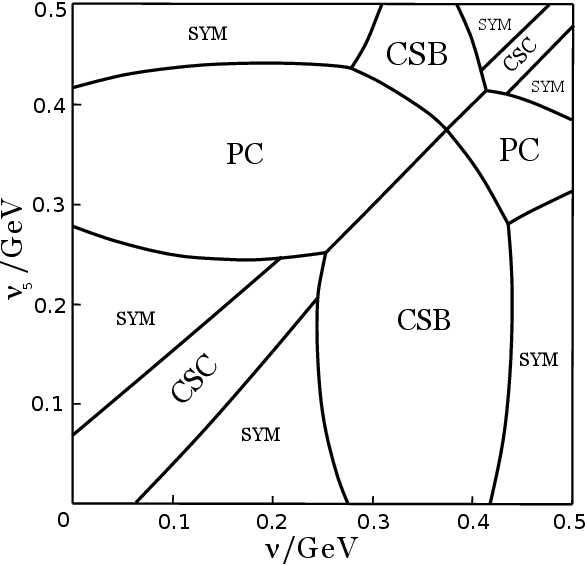}\\
%\label{fig1}
%\parbox[t]{0.45\textwidth}{
\caption{Left panel: $(\nu,\mu)$-phase diagram of the model at $\nu_5=0.3$ GeV, 
$\mu_5=0$ and $H=0.75G$. 
%}
 %}\hfill
%\parbox[t]{0.45\textwidth}{
%\caption{
Right panel: $(\nu,\nu_5)$-phase diagram of the model at $\mu=0.35$ GeV, $\mu_5=0.14$ GeV 
and $H=0.75G$. All the notations are the same as in Fig. 1.
 } %}
%\label{fig2}
\end{figure}

\begin{figure}
%----figure 3
\includegraphics[width=0.48\textwidth]{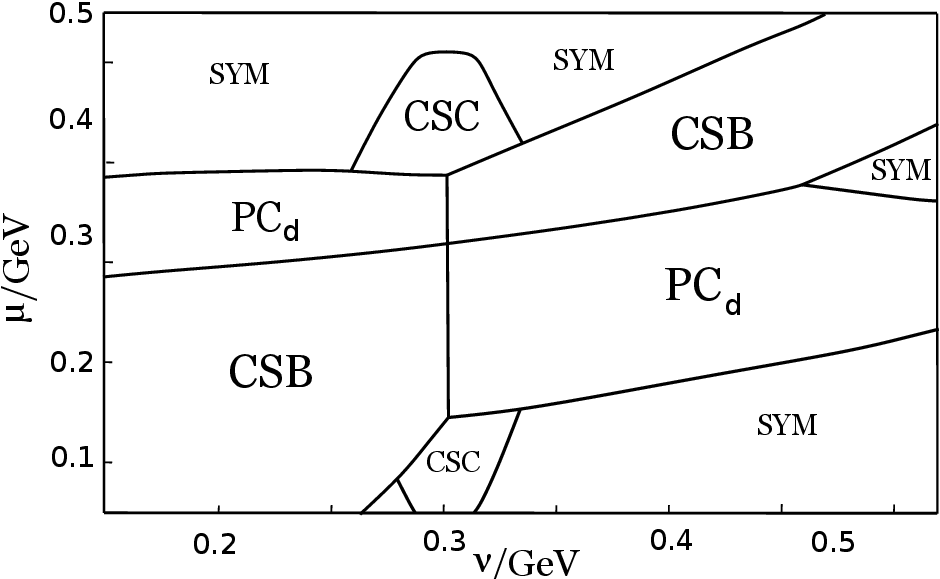}
 \hfill
\includegraphics[width=0.49\textwidth]{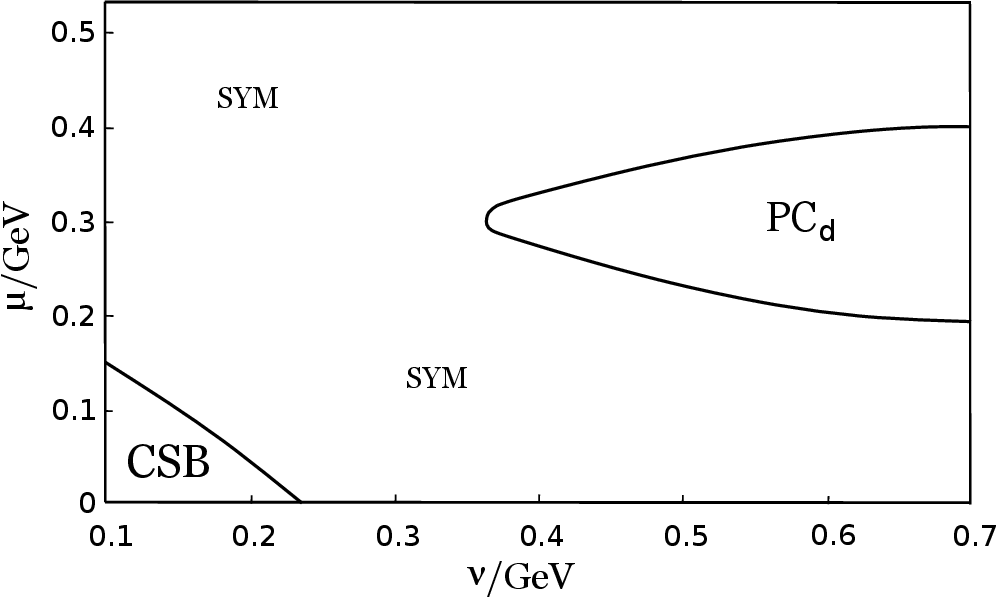}\\
%\label{fig1}
%\parbox[t]{0.45\textwidth}{
\caption{ $(\nu,\mu)$-phase diagrams of the model.  Left panel: The case 
of $\nu_5=0.3$ GeV and $\mu_5=-0.25$ GeV at $H=G$.
%}
 %}\hfill
%\parbox[t]{0.45\textwidth}{
%\caption{
Right panel:  The case of $\nu_5=0.3$ GeV and $\mu_5=0.2$ GeV at $H=0.75 G$. 
All the notations are the same as in Fig. 1. } %}
%\label{fig2}
\end{figure}

Let us discuss the phase diagram and charged pion condensation in chirally asymmetric dense quark matter in the 
light of color superconductivity phenomenon. Upon tedious investigations of the TDP 
(\ref{19}) (we study only the case $m=0$), it turned out that in most cases the charged PC$_d$ phase \footnote{To especially 
emphasize that we are dealing with the phenomenon of charged PC in medium with 
nonzero baryonic density, we will below use the abbreviation PC$_d$ of this phase in all 
phase portraits of the model.} is not affected by color superconductivity at 
all, as one can see in Fig. 1 (left) where the $(\mu_5,\mu)$-phase diagram at 
$\nu_5=-0.35$ GeV and $\nu=0$ is drawn. Note that all $(\mu_5,\mu)$-diagrams are very similar for various fixed
values of $\nu_5$ and $\nu$. And it is clear that in order for the charged  PC$_d$ phase to be 
generated, $|\nu_5|$ should be larger than $0.1$-$0.15$ GeV (see, e.g., in Figs. 2 and 3), 
does not matter positive or negative, otherwise the CSB phase would prevail in the phase 
structure in this case, but it is not the issue with color superconductivity, since the 
same is of course true in the framework of NJL model without diquark interaction
channel (see, e.g., in Refs. \cite{kkz18,kkz18-2}). 

One can see from different cross-sections of the full $(\mu,\nu,\mu_5,\nu_{5})$-phase diagram that in the rather extensive range of chemical  
potential space the CSC phase does not interfere with charged PC$_d$ phase at all (see in Figs. 2 (left) and 2 (right) or 3 (right)), where the ($\nu$, $\mu$)-phase diagrams  are presented. One can see in Fig. 2 (left) that the CSC phase resembles an oval shaped region, but it occupies a different from PC$_d$ phase region of chemical  
potential space. And does not meddle in charged pion condensation phenomenon.

Let us try to see when it is possible that CSC phase could in principle intervene in the 
charged PC phenomenon at nonzero baryon density. One can explore the cross-sections of the full 
phase diagram similar to the one in Fig. 1 (left) at various fixed values of $\nu$ and 
$\nu_5$ and behold, as we pointed out earlier, that the pictures are very similar. 
But one could note that in the case of $\nu\approx|\nu_5|$ there appears the band of CSC 
phase that crosses the PC$_d$ phase, as one could see in Fig. 1 (right). Nonetheless, one 
can see that in the region, where they interfere and compete, the charged PC$_d$ phase is 
always more energetically favorable and the shape and the size of charged PC$_d$ phase 
stays intact. Also let us note here that it is a rather peculiar choice of chemical 
potentials, very fine-tuned one. When $\nu=|\nu_5|$, the  CSB and charged PC$_d$ phases are 
almost degenerate, it is a region of the transition of these phases. And even in this 
specially chosen chemical potential case the CSC phase does not undermine the stability 
of the charged PC$_d$ phase in a bit. The whole charged PC$_d$ phase predicted in 
\cite{zhokhov1, zhokhov2,kkz18,kkz18-2,khunjua,khunjua2} previously, stays intact even 
in this case.

The fact that at $\nu\approx |\nu_5|$ there is  a region of phase transition between charged PC$_d$ and CSB phases could be easily illustrated by another cross-section of the full phase diagram, namely by the $(\nu,\nu_5)$ one depicted in Fig. 2 (right). Indeed, as it  can be easily seen in Fig. 2 (right), the band of CSC phase crosses the region of charged PC$_d$ phase in a rather small domain, where in addition the transition from CSB to charged PC$_d$ phase, and vise versa, occurs, i.e. in the region where these two phases are almost degenerate. 
It complies with Fig. 1 (right) where the same region is depicted but in another cross-section of the full $(\mu,nu,\mu_5,\nu_5)$-phase diagram of the model. So the main charged PC$_d$ phase region in Fig. 2 (right), the region at smaller $\nu$, lies beyond and is unbothered by the CSC phenomenon. 

One can see from another angle that in the rather extensive range of chemical  
potential space the CSC phase does not interfere with charged PC$_d$ phase at all.
And in those exceptional cases when the CSC and charged PC$_d$ phases could occupy the same region in the space of chemical  potentials, the charged PC$_d$ phase is still realized in the ground state of the system. This property of the phase structure of the model can be most successfully manifested in terms of its various $(\nu,\mu)$ cross-sections.
For example, contemplating the figures (see Figs. 3 (right) or 2 (left)) one could see that at $\mu_5>0$  CSC phase peacefully shares the phase diagram with charged PC$_d$ phase and they lie in different regions. (Remark that in Fig. 3 (right) CSC phase lies far beyond the region occupied by charged PC$_d$ phase and even does not present at the phase diagram. It lies in the region of $\mu>550$ MeV, i.e. outside the scope of the figure.) 
But one can find values of $\mu_5$ when these phases could go into clash with each other, e.g., for some values of $\mu_5<0$ the regions occupied by CSC and charged PC$_d$ phases can slightly overlap. The overlap region is still not extensive and these phases are elongated along different directions, i.e. axes with different chemical  
potentials. Having shown that there are such regions, it could be demonstrated that even if they overlap, what is more important, the charged PC$_d$ phase is nowhere undermined by CSC phase and turns out to be more energetically favorable in their intersection.
This property of the model is demonstrated by diagram in Fig. 3 (left). One should also notice that in Fig. 3 (left) in order to stress that it is a quite universal prediction, the value of $H$ was chosen to be $G$, and at this value we have the largest CSC phase and the biggest diquark condensate values (though this value of the  coupling constant $H$ cannot be now determined precisely, it is believed to be the largest physically motivated value for coupling constant of the quark-quark interaction channel). So in this case diquark condensation is the most favourable and CSC phase is the largest and it was chosen specially to show that even in this case the charged PC$_d$ phase survives in full extent and could be realized without any disturbance from CSC.

Let us briefly conclude this section: 
\begin{itemize}
\item 
Charged PC is generated by chiral imbalance %in a wide range of baryon densities 
and it is shown to be unbothered by color superconductivity phenomenon in the whole 
range of baryon density.
\item Charged PC in dense quark matter and color superconductivity phenomena as a rule 
occupies different regions of phase diagram, so one can say that they peacefully coexist.
\item 
Even if one fine-tunes the values of chemical 
potentials in order to reach the maximum crossing of charged PC$_d$ and CSC phases in 
the phase diagram and choose the largest possible coupling constant $H$ in the quark-quark 
interaction channel, the CSC phase does not substitute charged PC$_d$ phase in any region of phase 
diagram and charged PC$_d$ phase stays intact even in this specially chosen case.
\end{itemize}

\section{Summary and conclusions}

Pion condensation in dense quark matter has been investigated in several studies, 
including, e.g., Refs. \cite{son,he,ak,ekkz,kyll}. These studies demonstrated that 
charged pion condensation could arise in quark matter with nonzero baryon density and 
isospin imbalance.  
However, subsequent studies revealed that the region occupied by the charged PC$_d$ phase 
is negligibly small under real physical conditions (in the physical point, i. e. physical quark masses), and it vanishes 
completely when imposing beta-equilibrium and electrical neutrality conditions, 
as relevant to matter in the cores of neutron stars.

Subsequent investigations (see, e.g., in Refs. \cite{kkz18,kkz18-2}) revealed several conditions that could facilitate charged PC 
in dense quark matter. One of them, namely the chiral imbalance of dense quark medium, when 
$\mu_5\ne 0$ and/or $\mu_{I5}\ne 0$,
is a key factor promoting this phenomenon at a wide range of baryon chemical potentials. However, in these studies, the possibility of color superconductivity 
and quark-quark condensation was often disregarded for the sake of clarity and simplicity. 
This simplification was reasonable given the complexity of the problem, which required consideration of multiple condensates and chemical potentials.

It was initially assumed that charged PC$_d$ phase would not be entirely replaced by CSC 
one. However, there was no guarantee of this, and the survival of the charged CP$_d$ phase was 
primarily based on the argument that it encompasses a substantial portion of the phase 
diagram. But strictly speaking color superconductivity was neglected in the previous 
studies solely due to the simplicity reasons, and the prediction of charged PC$_d$ phase at 
nonzero baryon density in reality could be fully swept away by color superconductivity, 
especially at larger baryon densities. 
Hence, it was reasonable and crucial to revise the generation of charged pion condensation 
in chirally asymmetric dense quark matter, when the appearance of color superconductivity is also allowed in it.

In this paper, we investigate the phase structure of a generalized massless Nambu-Jona-Lasinio model (\ref{1}) in the presence of four chemical potentials: baryon $\mu_B=3\mu$, isospin $\mu_I=2\nu$, chiral isospin $\mu_{I5}=2\nu_5$, and chiral $\mu_5$ chemical potentials and at zero temperature. The model describes interactions between quarks in quark-antiquark and diquark channels and 
its primary objective is to provide a theoretical framework for studying the properties of dense quark matter. 

It has been shown that color superconductivity phenomenon, the phenomenon believed to be dominant at large baryon density, does not hinder the generation of charged pion condensation in dense quark matter by chiral imbalance. If $\mu_5\ne 0$ and/or $\mu_{I5}\ne 0$, then charged PC$_d$ phase in reality is never replaced by CSC phase even at large baryon density and even if one takes values of coupling constant $H$ in the whole range of physically motivated values.
Color superconductivity phenomenon and charged pion condensation in dense quark matter dwell in different regions of chemical  
potential space and almost do not interfere with each other. It is rather unexpected finding. It was expected that probably some regions occupied by charged PC$_d$ phase survive the inclusion of color superconductivity phenomenon in the consideration. But the expectation was that pion condensation phase would considerably dwindle and would be vastly replaced by CSC phase at least for rather large baryon density.
One can say that charged pion condensation in dense quark matter with chiral imbalance is shown to exist in a wide range of baryon densities and does not compete in a way with color superconductivity phenomenon. These two phenomena peacefully dwell in the QCD phase diagram and make it even more rich and interesting. It is shown in the framework of effective model, needless to say of course it would be great to probe this phenomenon from first principles but due to sign problem it is out of range of current capabilities of lattice QCD simulations. %It is an interesting phenomenon in context of moderate-energy heavy ion collisions experiments and cores of neutron stars and its mergers.
%%%%%%%%%%%%%%%%%%%

We expect that our results may shed light on the intricate phase structure of dense quark 
matter with isospin and chiral (isospin) imbalances, potentially providing valuable 
insights into the physics of intermediate-energy heavy ion collisions, as well as neutron 
star cores and its mergers. However, to make the predictions more realistic, we plan in the 
future to consider the problem taking into account the nonzero quark mass, as well as the 
electrical neutrality of the quark medium being in beta equilibrium.

%We anticipate 
%that our results might elucidate the intricate phase structure of dense quark matter with isospin and chiral (isospin) imbalances, potentially providing valuable insights into the physics of moderate-energy heavy ion collisions and the cores of neutron stars and their mergers.


\begin{thebibliography}{999}

\bibitem{njl}
Y. Nambu and G. Jona-Lasinio, Phys. Rev. {\bf 122}, 345 (1961); {\bf 124}, 246 (1961).

\bibitem{buballa2}
S.~P.~Klevansky,
  %``The Nambu-Jona-Lasinio model of quantum chromodynamics,''
  Rev.\ Mod.\ Phys.\  {\bf 64}, 649 (1992);
 % doi:10.1103/RevModPhys.64.649
  %%CITATION = doi:10.1103/RevModPhys.64.649;%%
D. Ebert, H. Reinhardt and M. K. Volkov, Prog. Part. Nucl. Phys. {\bf 33}, 1 (1994); 
T.~Inagaki, T.~Muta and S.~D.~Odintsov,
  %``Dynamical symmetry breaking in curved space-time: Four fermion interactions,''
  Prog.\ Theor.\ Phys.\ Suppl.\  {\bf 127}, 93 (1997);
 % doi:10.1143/PTPS.127.93   [hep-th/9711084].
  %%CITATION = doi:10.1143/PTPS.127.93;%%
%M. Buballa, Phys. Rep. {\bf 407}, 205 (2005); 
A. A. Garibli, R. G. Jafarov, and V. E. Rochev, 
Symmetry {\bf 11}, no. 5, 668 (2019).
%%CITATION = doi:10.3390/sym11050668;%%

\bibitem{buballa}
M. Buballa, Phys. Rep. {\bf 407}, 205 (2005).

\bibitem{zhokhov1}
T.~G.~Khunjua, K.~G.~Klimenko and R.~N.~Zhokhov,
 Symmetry {\bf 11}, no. 6, 778 (2019);
  %%CITATION = doi:10.3390/sym11060778;%%
  
\bibitem{zhokhov2}
T.~G.~Khunjua, K.~G.~Klimenko and R.~N.~Zhokhov, Particles {\bf 3}, no. 1, 62 (2020).
%%CITATION = doi:10.3390/particles3010006;%%  
  
\bibitem{son}
D. T.~Son and M. A.~Stephanov, Phys.\ Atom.\ Nucl.\  {\bf 64}, 834 (2001);
M. Loewe and C. Villavicencio, Phys. Rev. D {\bf 67}, 074034
(2003); M.~Frank, M.~Buballa and M.~Oertel,
%``Flavor mixing effects on the QCD phase diagram at nonvanishing isospin chemical potential: One or two phase transitions?,''
Phys. Lett. B \textbf{562}, 221-226 (2003);
%arXiv:1107.3859;
  %%CITATION = ARXIV:1107.3859;%%
D. C.~Duarte, R. L. S.~Farias and R. O.~Ramos,
  Phys.\ Rev.\  D {\bf 84}, 083525 (2011);
  %%CITATION = PHRVA,D84,083525;%%
D.~Ebert, K. G.~Klimenko, A. V.~Tyukov and V. C.~Zhukovsky,
%``Pion condensation of quark matter in the static Einstein %universe,''
  Eur.\ Phys.\ J.\ C {\bf 58}, 57 (2008).
  %%CITATION = ARXIV:0804.0765;%%

\bibitem{he}
L. He, M. Jin, and P. Zhuang, Phys. Rev. D {\bf 71}, 116001 (2005);
 %\bibitem{eklim}
D. Ebert and K. G. Klimenko, J.\ Phys.\ G {\bf 32}, 599 (2006);
%%CITATION = JPHGB,G32,599;%%
Eur.\ Phys.\ J.\  C {\bf 46}, 771 (2006);
%%CITATION = EPHJA,C46,771;%%
C.f.~Mu, L.y.~He and Y.x.~Liu,
  Phys.\ Rev.\  D {\bf 82}, 056006 (2010).
  %%CITATION = PHRVA,D82,056006;%%

  \bibitem{kyll}
H.~Abuki, R.~Anglani, R.~Gatto, M.~Pellicoro and M.~Ruggieri,
%``The Fate of pion condensation in quark matter: From the chiral to the real world,''
Phys. Rev. D \textbf{79}, 034032 (2009);
J. O.~Andersen and L.~Kyllingstad,
 J.\ Phys.\ G {\bf 37}, 015003 (2009).

\bibitem{ak}
J. O.~Andersen and T.~Brauner,
  Phys.\ Rev.\  D {\bf 78}, 014030 (2008);
  %%CITATION = PHRVA,D78,014030;%%
Y.~Jiang, K.~Ren, T.~Xia and P.~Zhuang,
  %``Meson Screening Mass in a Strongly Coupled Pion Superfluid,''
  Eur.\ Phys.\ J.\ C {\bf 71}, 1822 (2011);
  %%CITATION = doi:10.1140/epjc/s10052-011-1822-z;%%
  A.~Folkestad and J.~O.~Andersen,
  %``Thermodynamics and phase diagrams of Polyakov-loop extended chiral models,''
  Phys.\ Rev.\ D {\bf 99}, 054006 (2019);
 % doi:10.1103/PhysRevD.99.054006
 % [arXiv:1810.10573 [hep-ph]].
  %%CITATION = doi:10.1103/PhysRevD.99.054006;%%
   P.~Adhikari, J.~O.~Andersen and P.~Kneschke,
   Phys.\ Rev.\ D {\bf 98},  074016  (2018);
  %``QCD at finite isospin density: chiral perturbation theory confronts lattice data,''
   Eur.\ Phys.\ J.\ C {\bf 79},  874 (2019);
  %%CITATION = doi:10.1140/epjc/s10052-019-7381-4;%%
  J.~O.~Andersen, P.~Adhikari and P.~Kneschke,
%``Pion condensation and QCD phase diagram at finite isospin density,''
arXiv:1810.00419 [hep-ph].

\bibitem{ekkz}
 D.~Ebert, T. G.~Khunjua, K. G.~Klimenko and V. C.~Zhukovsky,
  %``Charged pion condensation phenomenon of dense baryonic matter induced by finite volume: The NJL(2) model consideration,''
  Int.\ J.\ Mod.\ Phys.\ A {\bf 27}, 1250162 (2012);
  %%CITATION = doi:10.1142/S0217751X1250162X;%%
  %\bibitem{gkkz}
  N. V.~Gubina, K. G.~Klimenko, S. G.~Kurbanov and V. C.~Zhukovsky,
  Phys.\ Rev.\ D {\bf 86}, 085011 (2012).
  %%CITATION = doi:10.1103/PhysRevD.86.085011;%%

  \bibitem{Mammarella:2015pxa} 
  A.~Mammarella and M.~Mannarelli,
  %``Intriguing aspects of meson condensation,''
  Phys.\ Rev.\ D {\bf 92},  085025 (2015);
  %\bibitem{Carignano:2016lxe} 
S.~Carignano, L.~Lepori, A.~Mammarella, M.~Mannarelli and G.~Pagliaroli,
  %``Scrutinizing the pion condensed phase,''
  Eur.\ Phys.\ J.\ A {\bf 53},  35 (2017);
 % doi:10.1140/epja/i2017-12221-x
 % [arXiv:1610.06097 [hep-ph]].
  %%CITATION = doi:10.1140/epja/i2017-12221-x;%%
  %arXiv:1610.06097 [hep-ph].
M.~Mannarelli,
  %``Meson Condensation,''
  Particles {\bf 2}, no. 3, 411 (2019).
  %doi:10.3390/particles2030025
  %[arXiv:1908.02042 [hep-ph]].
  %%CITATION = doi:10.3390/particles2030025;%%  
  
\bibitem{Andersen:2018nzq}
J.~O.~Andersen and P.~Kneschke,
       %``Bose-Einstein condensation and pion stars,''
arXiv:1807.08951 [hep-ph];
B.~B.~Brandt, G.~Endrodi, E.~S.~Fraga, M.~Hippert, J.~Schaffner-Bielich and S.~Schmalzbauer,
Phys.\ Rev.\ D {\bf 98},  094510 (2018).
       
\bibitem{Ayala}
A.~Ayala, B.~S.~Lopes, R.~L.~S.~Farias and L.~C.~Parra,
%``Describing the speed of sound peak of isospin-asymmetric cold strongly interacting matter using effective models,''
arXiv:2310.13130 [hep-ph].

  \bibitem{andrianov}
 %A. A.~Andrianov, D.~Espriu and X.~Planells,
  %``An effective QCD Lagrangian in the presence of an axial chemical potential,''
  %Eur.\ Phys.\ J.\ C {\bf 73}, 2294 (2013);
 % %%CITATION = ARXIV:1210.7712;%%
 %Eur.\ Phys.\ J.\ C {\bf 74}, 2776 (2014);
 % %%CITATION = ARXIV:1310.4416;%%
R.~Gatto and M.~Ruggieri,
  %``Hot Quark Matter with an Axial Chemical Potential,''
  Phys.\ Rev.\ D {\bf 85}, 054013 (2012);
  %%CITATION = ARXIV:1110.4904;%%
 L.~Yu, H.~Liu and M.~Huang,
  %``Spontaneous generation of local CP violation and inverse magnetic catalysis,''
 Phys.\ Rev.\ D {\bf 90}, 074009 (2014);
  %%CITATION = doi:10.1103/PhysRevD.90.074009;%%
%L.~Yu, H.~Liu and M.~Huang,
 Phys.\ Rev.\ D {\bf 94}, 014026 (2016);
  %%CITATION = ARXIV:1511.03073;%%
 M.~Ruggieri and G.~X.~Peng,
  %``Critical Temperature of Chiral Symmetry Restoration for Quark Matter with a Chiral Chemical Potential,''
  J.\ Phys.\ G {\bf 43}, no. 12, 125101 (2016);
  %doi:10.1088/0954-3899/43/12/125101
  %[arXiv:1602.05250 [hep-ph]].
  %%CITATION = doi:10.1088/0954-3899/43/12/125101;%%
%\bibitem{Andrianov:2019fwz}
  A.~A.~Andrianov, V.~A.~Andrianov and D.~Espriu,
  %``Chiral perturbation theory vs. Linear Sigma Model in a chiral imbalance medium,''
  Particles {\bf 3}, no. 1, 15 (2020);
 % doi:10.3390/particles3010002
 % [arXiv:1908.09118 [hep-th]].
  %%CITATION = doi:10.3390/particles3010002;%%
D.~Espriu, A.~G.~Nicola and A.~Vioque-Rodríguez,
  %``Chiral perturbation theory for nonzero chiral imbalance,''
  arXiv:2002.11696 [hep-ph].
  %%CITATION = ARXIV:2002.11696;%%

\bibitem{cao}
G.~Cao and P.~Zhuang,
  %``Effects of chiral imbalance and magnetic field on pion superfluidity and color superconductivity,''
  Phys.\ Rev.\ D {\bf 92}, 105030 (2015).
  %%CITATION = doi:10.1103/PhysRevD.92.105030;%%

\bibitem{braguta}
V. V.~Braguta and A. Y.~Kotov,
 Phys.\ Rev.\ D {\bf 93}, 105025 (2016).
  %%CITATION = doi:10.1103/PhysRevD.93.105025;%%

%\bibitem{braguta3}  
%V.~V.~Braguta, V.~A.~Goy, E.-M.~Ilgenfritz, A.~Y.~Kotov, A.~V.~Molochkov, M.~Muller-Preussker and B.~Petersson,
 %``Two-Color QCD with Non-zero Chiral Chemical Potential,''
% JHEP {\bf 1506}, 094 (2015).
 
\bibitem{braguta2}  
V.~V.~Braguta, E.~M.~Ilgenfritz, A.~Y.~Kotov, B.~Petersson and S.~A.~Skinderev,
%``Study of QCD Phase Diagram with Non-Zero Chiral Chemical Potential,''
Phys.\ Rev.\ D {\bf 93},  034509  (2016).
%[arXiv:1512.05873 [hep-lat]];
  
\bibitem{Farias}
R.~L.~S.~Farias, D.~C.~Duarte, G.~Krein and R.~O.~Ramos,
%``Thermodynamics of quark matter with a chiral imbalance,''
Phys. Rev. D \textbf{94}, no.7, 074011 (2016).  
       
\bibitem{Khunjua:2019lbv}
T.~G.~Khunjua, K.~G.~Klimenko and R.~N.~Zhokhov,
%``Dualities and inhomogeneous phases in dense quark matter with chiral and isospin imbalances in the framework of effective model,''
JHEP \textbf{06}, 006 (2019).

\bibitem{kkz18}
T.~G.~Khunjua, K.~G.~Klimenko and R.~N.~Zhokhov,
%``Dense baryon matter with isospin and chiral imbalance in the framework of NJL$_4$ model at large $N_c$: duality between chiral symmetry breaking and charged pion condensation,''
Phys. Rev. D \textbf{98}, no.5, 054030 (2018).
%doi:10.1103/PhysRevD.98.054030
%[arXiv:1804.01014 [hep-ph]].
%doi:10.1103/PhysRevD.97.054036
%[arXiv:1710.09706 [hep-ph]].

\bibitem{kkz18-2}
T.~G.~Khunjua, K.~G.~Klimenko and R.~N.~Zhokhov,
%``Dense baryon matter with isospin and chiral imbalance in the framework of NJL$_4$ model at large $N_c$: duality between chiral symmetry breaking and charged pion condensation,''
Phys. Rev. D \textbf{97}, no.5, 054036 (2018).

\bibitem{khunjua}
T.~G.~Khunjua, K.~G.~Klimenko and R.~N.~Zhokhov,
%``Electrical neutrality and $\beta$-equilibrium conditions in dense quark matter: generation of charged pion condensation by chiral imbalance,''
Phys. Rev. D \textbf{100}, no.3, 034009 (2019); Moscow Univ. Phys. Bull. \textbf{74}, no.5, 473 (2019);  
Acta Phys. Polon. Supp. \textbf{14}, 67 (2021).

\bibitem{khunjua2}
T.~G.~Khunjua, K.~G.~Klimenko and R.~N.~Zhokhov,
%``Chiral imbalanced hot and dense quark matter: NJL analysis at the physical point and comparison with lattice QCD,''
Eur. Phys. J. C \textbf{79}, no.2, 151 (2019).

\bibitem{khkz}
T.~G.~Khunjua, K.~G.~Klimenko and R.~N.~Zhokhov,
Eur. Phys. J. C \textbf{80}, no.10, 995 (2020). 

%\bibitem{Khunjua:2018dbm}
%T.~G.~Khunjua, K.~G.~Klimenko and R.~N.~Zhokhov,
%``QCD phase diagram with chiral imbalance in NJL model: duality and lattice QCD results,''
%J. Phys. Conf. Ser. \textbf{1390}, no.1, 012015 (2019)

\bibitem{alford}
I. A. Shovkovy, Found. Phys. {\bf 35}, 1309 (2005);
 M.~Huang,
%``Color superconductivity at moderate baryon density,''
Int.\ J.\ Mod.\ Phys.\ E {\bf 14}, 675 (2005);
 % doi:10.1142/S0218301305003491
 % [hep-ph/0409167].
%%CITATION = doi:10.1142/S0218301305003491;%%
 K.~G.~Klimenko and D.~Ebert,
%``Mesons and diquarks in a dense quark medium with color superconductivity,''
Theor.\ Math.\ Phys.\  {\bf 150}, 82 (2007) [Teor.\ Mat.\ Fiz.\  {\bf 150}, 95 (2007)];
%doi:10.1007/s11232-007-0006-3
%%CITATION = doi:10.1007/s11232-007-0006-3;%%
M. G.~Alford, A.~Schmitt, K.~Rajagopal, and T.~Sch\"afer,
 Rev.\ Mod.\ Phys\  {\bf 80}, 1455 (2008);
  %%CITATION = RMPHA,80,1455;%%
 E.~J.~Ferrer and V.~de la Incera,
  %``Magnetism in Dense Quark Matter,''
  Lect.\ Notes Phys.\  {\bf 871}, 399 (2013).
 % doi:10.1007/978-3-642-37305-3_16
 % [arXiv:1208.5179 [nucl-th]].
  %%CITATION = doi:10.1007/978-3-642-37305-3_16;%%  

\bibitem{CSC}
T.~G.~Khunjua, K.~G.~Klimenko and R.~N.~Zhokhov,
%``Dual properties of dense quark matter with color superconductivity phenomenon,''
Phys. Rev. D \textbf{108}, no.12, 125011 (2023)
%doi:10.1103/PhysRevD.108.125011
[arXiv:2310.08211 [hep-ph]].

\bibitem{pionnuclear}
A. B. Migdal, Zh. Eksp. Teor. Fiz. {\bf 61}, 2210  (1971) [Sov. Phys. JETP {\bf 36}, 1052 (1973)].

\bibitem{pionnuclearVoskresensky}
 A. B. Migdal, E. E. Saperstein, M. A. Troitsky and D. N. Voskresensky, Phys. Rept. 
{\bf 192}, 179  (1990);
 D. G. Yakovlev, K. P. Levenfish and Y. A. Shibanov, Phys. Usp. {\bf 42}, 737  (1999) 
 [astro-ph/9906456];
%[28] T. Tatsumi, Prog. Theor. Phys. 63 (1980) 1252; Prog. Theor. Phys. 68 (1982) 1231; Prog. Theor. Phys. 69 (1983) 1137.
%[29] 
T. Takatsuka and R. Tamagaki, 
Prog. Theor. Phys. Suppl. {\bf 112}, 107  (1993); %Prog. Theor. Phys. 97 (1997) 263; 
T. Takatsuka, R. Tamagaki and T. Tatsumi, Prog. Theor. Phys. Suppl. {\bf 112}, 67 (1993).

\bibitem{Yasuda}
J.~Yasuda, M.~Sasano, R.~G.~T.~Zegers, \textit{et al.}
%``Extraction of the Landau-Migdal Parameter from the Gamow-Teller Giant Resonance in Sn132,''
Phys. Rev. Lett. \textbf{121}, no.13, 132501 (2018).

\bibitem{VoskresenskyPion}
D. N. Voskresensky
%Pion Softening and Pion Condensation. 
Phys. Atom. Nuclei {\bf 83}, 188 (2020). %https://doi.org/10.1134/S1063778820020301

\bibitem{Voskresensky:2022gts}
D.~N.~Voskresensky,
%``S-wave pion condensation in symmetric nuclear matter,''
Phys. Rev. D \textbf{105}, no.11, 116007 (2022).

%\cite{Dohi:2021lbu}
\bibitem{Dohi:2021lbu}
A.~Dohi, H.~Liu, T.~Noda and M.~a.~Hashimoto,
%``Cooling of isolated neutron stars with pion condensation: Possible fast cooling in a low-symmetry energy model,''
Int. J. Mod. Phys. E \textbf{31}, no.02, 2250006 (2022).

\bibitem{Ohnishi:2008ng}
A.~Ohnishi, D.~Jido, T.~Sekihara and K.~Tsubakihara,
%``Possibility of s-wave pion condensates in neutron stars revisited,''
Phys. Rev. C \textbf{80}, 038202 (2009)
[arXiv:0810.3531 [nucl-th]].

\bibitem{EricsonWeise}
T.E.O. Ericson and W. Weise, Pions and Nuclei (Oxford University Press, Oxford, 1988);
T. Wakasa et al., Phys. Rev. C {\bf 55}, 2909 (1997); T. Suzuki and H. Sakai, Phys. Lett. B {\bf 455}, 25 (1999).

\bibitem{Kuznietsov:2021lax}
V.~A.~Kuznietsov, O.~S.~Stashko, O.~V.~Savchuk and M.~I.~Gorenstein,
%``Critical point and Bose-Einstein condensation in pion matter,''
Phys. Rev. C \textbf{104}, no.5, 055202 (2021)
[arXiv:2108.08140 [hep-ph]].
%5 citations counted in INSPIRE as of 03 Jul 2023

\bibitem{Stashko:2023gnn}
O.~S.~Stashko, O.~V.~Savchuk, L.~M.~Satarov, I.~N.~Mishustin, M.~I.~Gorenstein and V.~I.~Zhdanov,
%``Pion stars embedded in neutrino clouds,''
Phys. Rev. D \textbf{107}, no.11, 114025 (2023)
[arXiv:2303.06190 [hep-ph]].
%0 citations counted in INSPIRE as of 03 Jul 2023

%\cite{Brandt:2018bwq}
\bibitem{Brandt:2018bwq}
B.~B.~Brandt, G.~Endrodi, E.~S.~Fraga, M.~Hippert, J.~Schaffner-Bielich and S.~Schmalzbauer,
%``New class of compact stars: Pion stars,''
Phys. Rev. D \textbf{98}, no.9, 094510  (2018)
[arXiv:1802.06685 [hep-ph]].
%63 citations counted in INSPIRE as of 03 Jul 2023

\bibitem{Vijayan:2023qrt}
V.~Vijayan, N.~Rahman, A.~Bauswein, G.~Mart\'\i{}nez-Pinedo and I.~L.~Arbina,
%``Impact of pions on binary neutron star merger,''
Phys. Rev. D \textbf{108}, no.2, 023020 (2023)
[arXiv:2302.12055 [astro-ph.HE]].

\bibitem{Vovchenko:2020crk}
V.~Vovchenko, B.~B.~Brandt, F.~Cuteri, G.~Endr\H{o}di, F.~Hajkarim and J.~Schaffner-Bielich,
Phys. Rev. Lett. \textbf{126}, no.1, 012701  (2021)
[arXiv:2009.02309 [hep-ph]].
%29 citations counted in INSPIRE as of 03 Jul 2023

\bibitem{Cao:2021gfk}
G.~Cao, L.~He and P.~Zhang,
%``Reentrant pion superfluidity and cosmic trajectories within a PNJL model,''
Phys. Rev. D \textbf{104}, no.5, 054007  (2021)
[arXiv:2105.08932 [hep-ph]].

\bibitem{Wygas:2018otj}
M.~M.~Wygas, I.~M.~Oldengott, D.~B\"odeker and D.~J.~Schwarz,
%``Cosmic QCD Epoch at Nonvanishing Lepton Asymmetry,''
Phys. Rev. Lett. \textbf{121}, no.20, 201302  (2018)
[arXiv:1807.10815 [hep-ph]].

\bibitem{Liu:2017spl}
Y.~Liu and I.~Zahed,
%``Pion Condensation by Rotation in a Magnetic field,''
Phys. Rev. Lett. \textbf{120}, no.3, 032001  (2018)
[arXiv:1711.08354 [hep-ph]].

\bibitem{Cao:2019ctl}
G.~Cao and L.~He,
Phys. Rev. D \textbf{100}, no.9, 094015  (2019)
[arXiv:1910.02728 [nucl-th]].

\bibitem{Ebert}
D.~Ebert, K.~G.~Klimenko, V.~C.~Zhukovsky and A.~M.~Fedotov,
Eur. Phys. J. C \textbf{49}, 709 (2007).

\bibitem{2color}
T.~G.~Khunjua, K.~G.~Klimenko and R.~N.~Zhokhov,
%``Influence of chiral chemical potential \ensuremath{\mu}5 on phase structure of the two-color quark matter,''
Phys. Rev. D \textbf{106}, no.4, 045008 (2022).

 
\end{thebibliography}
\end{document}